\documentclass[aps,prl,reprint,superscriptaddress,showpacs]{revtex4-1}

\usepackage{graphicx}
\usepackage{dcolumn}
\usepackage{bm}
\usepackage{amsmath}
\usepackage{amssymb}
\usepackage{xcolor}
\usepackage{subfigure}
\usepackage{array}

\begin{document}

\preprint{To be submitted to PRL}

\title{Frozen Patterns of Impacted Droplets: From Conical Tips to Toroidal Shapes}

\author{Man Hu}
\thanks{Contributed equally}
\author{Feng Wang}
\thanks{Contributed equally}
\author{Qian Tao}
\author{Li Chen}
\affiliation{Department of Aeronautics and Astronautics, Fudan University, Shanghai, 200433, China}
\author{Shmuel M. Rubinstein}
\affiliation{John A. Paulson School of Engineering and Applied Sciences, Harvard University, Cambridge,
Massachusetts 02138, USA}
\author{Daosheng Deng}
\email{dsdeng@fudan.edu.cn}
\affiliation{Department of Aeronautics and Astronautics, Fudan University, Shanghai, 200433, China}


\date{\today}

\begin{abstract}
We report frozen patterns for the water droplets impacting on a cold substrate through fast-speed images. These patterns can be manipulated by several physical parameters (the droplet size, falling height, and substrate temperature), and the scaling analysis has a remarkable agreement with the phase diagram. The observed double-concentric toroidal shape is attributed to the correlation between the impacting dynamics and freezing process, as confirmed by the spatiotemporal evolution of the droplet temperature, the identified timescale associated with the morphology and solidification ($t_{inn}\simeq \tau_{sol}$), and the ice front-advection model. These results for frozen patterns provide insight into the complex interplay of the rapid impacting hydrodynamics, the transient heat transfer, and the intricate solidification process.

\end{abstract}

\maketitle


Over the past decade, the elegant and beautiful splashing patterns during water droplets impacting on a solid substrate at room temperature have been extensively investigated   \cite{yarin2006drop,josserand2016drop,xu2005drop,kolinski2012skating}. Recently, the study of droplets impacting on a cold substrate has emerged, and many fascinating and intriguing physical phenomena have been revealed \cite{marin2014universality,ghabache2016frozen,ruiter2018selfpeeling,kant2020fast}. For example, the universal ice-cone formation of freezing droplets is characterized by a tip singularity due to the geometric confinement of the freezing fronts \cite{marin2014universality}. As the substrate temperature is sufficiently low, the crack pattern is observed in the form of fragmentation or hierarchical fracture \cite{ghabache2016frozen}. By controlling the thermal properties of the substrate, the self-peeling of impacting droplets occurs \cite{ruiter2018selfpeeling}.  A peculiar freezing morphology at the liquid-solid interface, as explored by the total internal reflection technique, is explained by an ice front-advection model due to the rapid growth of crystals and the sequential advection by internal viscous flow \cite{kant2020fast}. These scientific advancements have the promising implications for the relevant technologies, such as icing and anti-icing \cite{cebeci2003aircraft,vasileiou2017imparting}, or solidification for liquid metals and 3D printing \cite{davis2001theory,aziz2000impact,haferl2003experimental,vaezi2013review,gielen2020solidification}.

In this paper, we report frozen patterns for the droplets falling from various heights on a cold substrate through high-speed images. These patterns can be controlled by several physical parameters (the droplet size, falling height, and the substrate temperature); and their transition can be remarkably described by a simple power law from the scaling analysis. The observed double-concentric toroidal shape is attributed to the strong correlation between the impacting hydrodynamics and solidification process, as confirmed by the spatiotemporal evolution of the droplet temperature, the identified timescale associated with the morphology and solidification ($t_{inn}\simeq \tau_{sol}$), and the ice front-advection model.

\begin{figure*}[!t]{}
    \center
		  \includegraphics[scale=0.6]{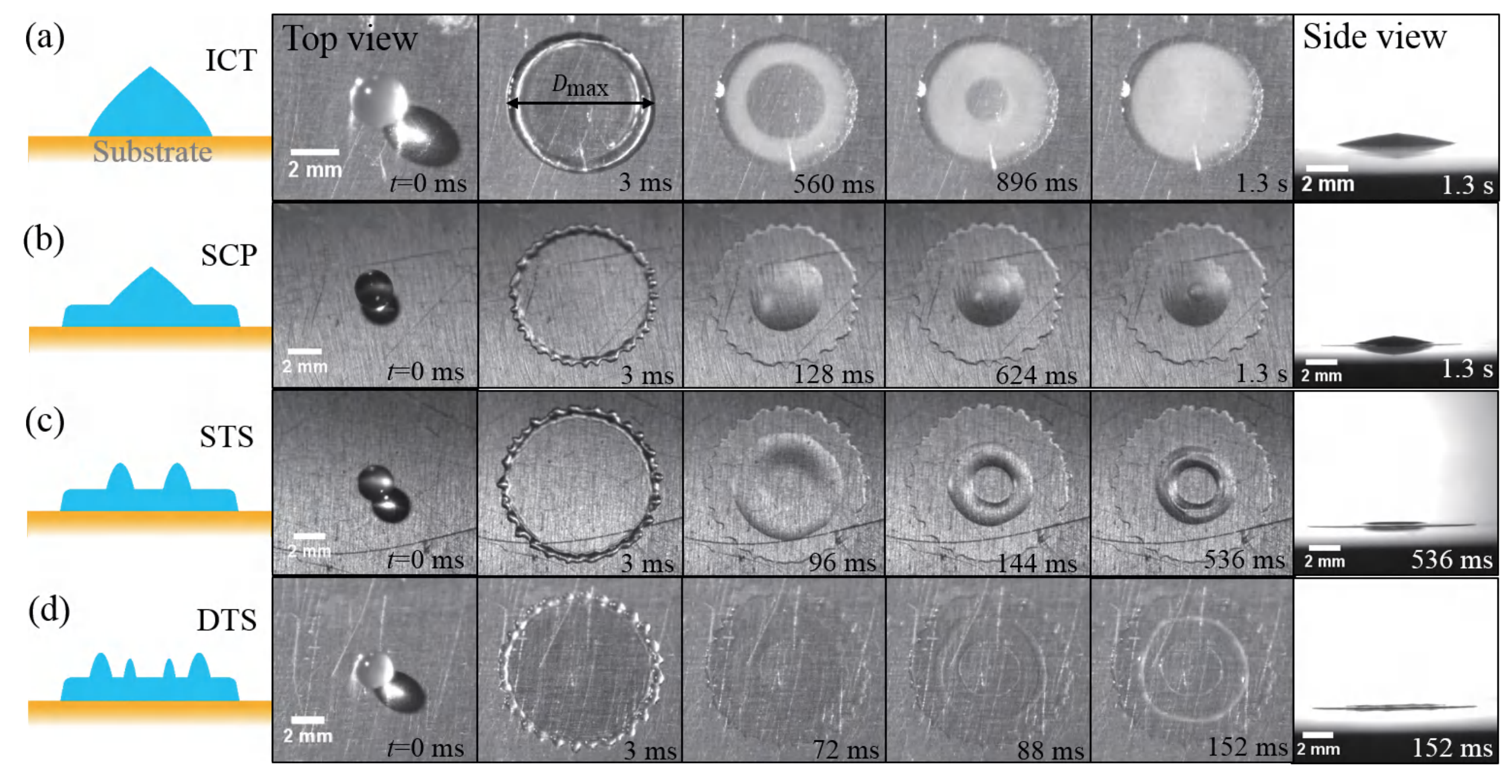}
    \caption{Four frozen patterns are visualized by the high-speed images for the spreading dynamics and the subsequent solidification process during the droplet impacted on a cold substrate at $T_s = -15 \, \mathrm{^{\circ} C}$. (a) Ice with a conical tip, (b) a spherical cap on the pancake, (c) a single toroidal shape, and (d) a double-concentric toroidal shape for $H_0$ = 10, 35, 70, and 80 cm, respectively. $t = 0$ ms is defined as droplets landing on or starting to impact on the substrate, as shown by the first-column images.}
                \label{fig_1}
\end{figure*}

The experiments were performed at room temperature, and the substrate temperature ($T_s <0\, \mathrm{^{\circ}C}$) was controlled by partially immersing the aluminium plate into liquid nitrogen. The outlet of the micrometer syringe (Gillmont GS-1200, Cole-Parmer Inc, 2 ml) is connected with a needle, the inner diameter of which controls the droplet size (volume $V_0$ or diameter $D_0 = 2R _0$). Then the distilled water droplet was released to impact on the cold substrate (SM).

The impacting dynamics is captured by fast-speed cameras (Phantom V611) from the top view and the side view (Fig. 1), and four types of frozen patterns are identified for the droplets ($D_0 = 2R _0 \approx 2.4$ mm) released at various heights $H_0$ = 10, 35, 70 and 80 cm ($T_s = - 15\mathrm{^{\circ} C}$) [Video 1-4 in supplementary materials(SM)].  First, for the lowest $H_0$ = 10 cm (Fig. 1a), the droplet impacts on the cold substrate ($t =0$ ms) and spreads outward radially until reaching its maximum diameter ($D_{max}$) at $t = 3$ ms. Since the ice/water front moves radially towards its center during solidification, the droplet is gradually frozen into ice,  while a pointy tip (side view) is formed \cite{marin2014universality}. Eventually, the final frozen pattern is the ice with a conical tip (ICT).

Secondly, for the increased $H_0$ = 35 cm (Fig. 1b), after spreading to $D_{max}$, its spreading edge is pinned to subject the regime of the stationary contact line rather than recoiling immediately   \cite{thievenaz2019solidification}. During the stationary contact line period ($\tau_{SCL}$), the underneath liquid is directly contacted with the cold substrate, resulting in the growth of the bottom-pancake ice. After the stationary contact line stage, the remaining top-layer liquid with larger aspect ratio is unstable due to surface tension, and retracts to form a spherical cap, while the pointy tip still appears at the final freezing stage (side view). Then, the frozen pattern \cite{thievenaz2019solidification} is a spherical cap on the pancake (SCP).

However, at much higher $H_0$, more intriguing final frozen patterns are observed that the distinctive toroidal shapes together with the inner holes (or the inner valley regions), as clearly visible in Fig. 1c and d.  For $H_0$ = 70 cm (Fig. 1c), the frozen pattern is exemplified by a single toroidal shape (STS), reminiscent of a doughnut shape during the impacted droplets at room temperature \cite{renardy2003pyramidal} or during laser-induced forward transfer \cite{visser2015toward}. For $H_0$ = 80 cm (Fig. 1d), the frozen pattern is represented by an intriguing double-concentric toroidal shape (DTS), which is observed here for the first time, to the best of our knowledge.

\begin{figure*}[!t]
    \center
    \includegraphics[scale=0.55]{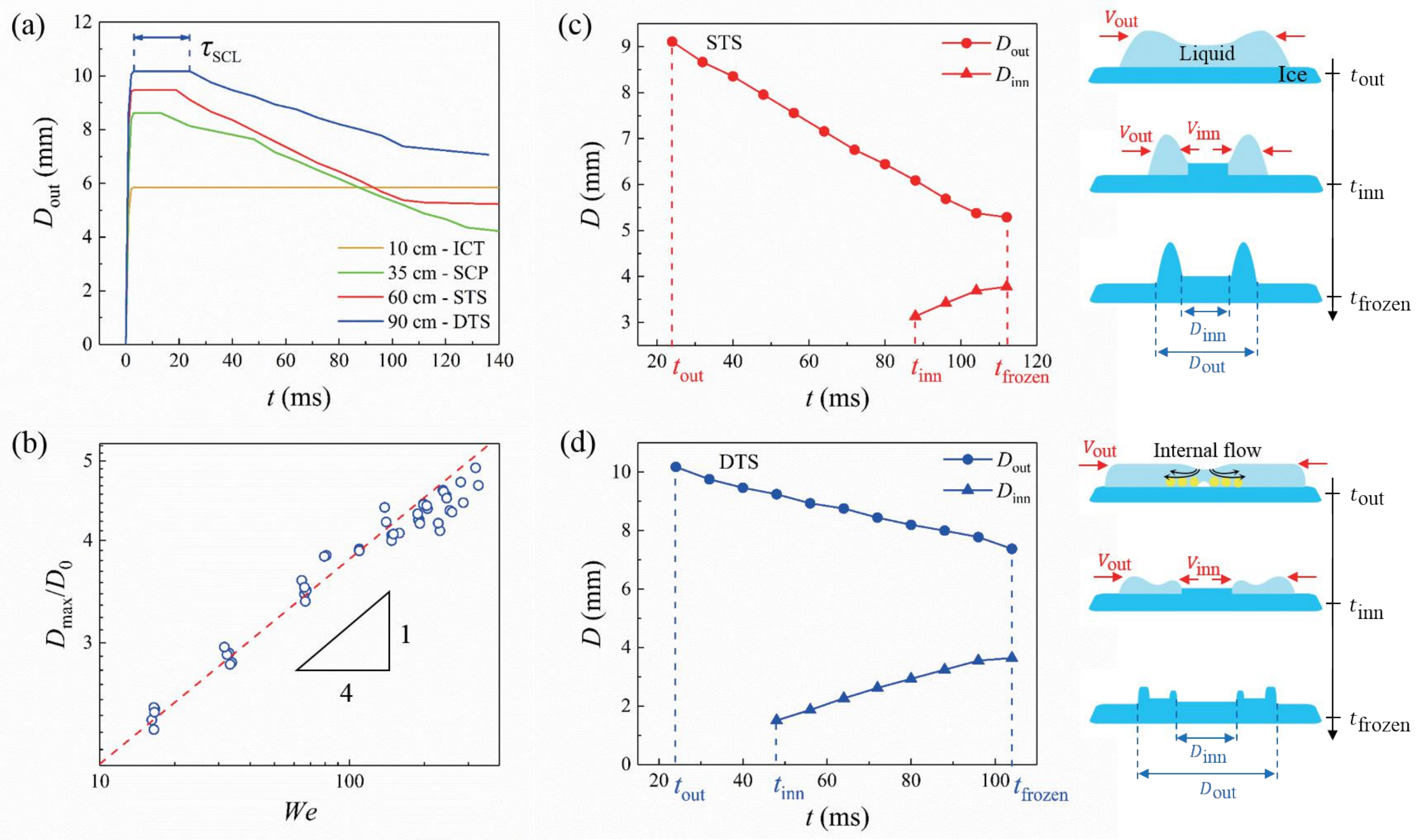}
    \caption{Formation of toroidal shapes at $T_s = -15 \, \mathrm{^{\circ} C}$. (a) The full dynamics of $D_{out}$, indicating $\tau_{SCL}$ during which the bottom-pancake ice is formed. (b) The scaling law of $D_{max} \sim We^{1/4}$. (c, d) Dynamics of $D_{out}$ and $D_{inn}$ for STS and DTS. Sketch mechanism for STS due to the doughnut shape at $t_{out}$, and for DTS based on the ice front-advection model.}
                \label{fig_2}
\end{figure*}
\begin{table}[b]
\centering
\caption{\label{comparison} Velocity of the rim boundary of the top-layer liquid film (mm/s)}
\begin{tabular}{p{1.6cm}<{\centering}p{1.6cm}<{\centering}p{1.6cm}<{\centering}p{0.05cm}<{\centering}p{1.6cm}<{\centering}p{1.6cm}<{\centering}}
\hline
\hline
    &\multicolumn{2}{c}{STS $(H_0 = 60$ cm)} & & \multicolumn{2}{c}{DTS $(H_0 = 90$ cm)}\\ \hline
    & $v_{out} $   & $v_{inn}$    & & $v_{out} $   & $v_{inn} $    \\ \hline
Experiment & $23.44\pm 0.95$ & $18.79\pm 3.25$  &  & $16.28\pm 0.11$ & $22.96\pm 0.67$ \\
Theory & $21.3$ & $13.4$ & & $20.5$ & $13.4$ \\ \hline \hline
\end{tabular}
\label{table:velocity}
\end{table}
In order to understand the observed toroidal shapes, the dynamics of the outer rim boundary of the top-layer liquid film (its diameter $D_{out}$) is compared (Fig. 2a). Similar to SCP \cite{thievenaz2019solidification}, the stage of the stationary contact line still exists for STS and DTS to form the bottom-pancake ice, and  $\tau_{SCL}=10^{5}\eta (h-70)/\gamma$  ($\gamma, \eta$ for the surface tension and viscosity of water), where the initial liquid thickness $h$ ($\mathrm{\mu}$m scale here) is expressed as below,
 \begin{equation}\label{equthickness}
 h = 4V_{0}/\pi D_{max}^{2}.
 \end{equation}
The calculated $\tau_{SCL} \sim 23$ ms is comparable with the experimental $\tau_{SCL}$ ( 16 ms for STS and 21 ms for DTS).

Despite the complicated frozen dynamics for these patterns, similar to the drop impacting on solid surface at room temperature \cite{clanet2004maximal}, the following scaling law for $D_{max}$ holds well (Fig. 2b), indicating that the spreading process is dominated by the impact dynamics,
\begin{equation}\label{Dmaxscaling}
  D_{max} \sim D_0 We^{1/4},
\end{equation}
where $\emph{We} = \rho R_{0}U_{0}^{2}/\gamma = 2\rho gH_{0}R_{0}/\gamma$ is Weber number ($\rho$ for the density of water, and $U_0$ for the impacted velocity).

For STS and DTS, the dynamics of the inner rim boundary of the top-layer liquid (its diameter $D_{inn}$) is further demonstrated (Fig. 2c and d). At $t_{out}$ after $\tau_{SCL}$, the spreading edge of the top-layer liquid is unpinned and begins to recoil, resulting in the decreased $D_{out}$ with a constant velocity $v_{out}$ (that is attained by the linear fitting of $D_{out}$ in experiments), as shown in Table \ref{table:velocity}.  Afterwards at $t_{inn}$, the inner hole or the inner rim boundary of the top-layer liquid occurs or is noticeably observed from the high-speed images, and $D_{inn}$ expands radially outwards with a constant velocity $v_{inn}$ (that is attained by the linear fitting of $D_{inn}$ in experiments). Then at $t_{frozen}$, the droplets are completely frozen into the final patterns.

Quantitatively, the experimental velocity of $v_{out}$ and $v_{inn}$ for both STS and DTS is reasonably consistent with the theoretical estimation. From the characteristic length scale ($L \sim 0.5 D_{max}$) and capillary timescale [$\tau_{cap}=(4\pi \rho D_{max}^{3}/3\gamma)^{1/2}$], the obtained theoretical $v_{out}= L/\tau_{cap} \approx  20$ mm/s is comparable with the experimental $v_{out}$, implying that the capillary force plays an important role for the top-layer liquid film retraction. For $v_{inn}$, similar to the dewetting rim model for the liquid film with an initial dry hole \cite{degennes2013capillarity,snoeijer2010asymptotic}, $v_{inn}=\gamma \theta_{r}^{3}/6\eta\textup{ln}$, where $\theta_{r} \sim 0.1\,\mathrm{rad}$ is the receding contact angle of water on ice \cite{watericeangle}, and $\textup{ln} \sim 1$ for a logarithmic factor \cite{degennes2013capillarity}. Then again, the theoretical $v_{inn}\approx 13$ mm/s fairly agrees with the experimental values. Additionally, this dynamics of DTS and the frozen double-concentric toroidal shapes can be reproduced well (SM).

Physically, STS (sketch in Fig. 2c) is reminiscent of the toroidal shape for a droplet impacting on a superhydrophobic solid surface at room temperature \cite{renardy2003pyramidal}. The faster spreading of the droplet falling at a larger $H_0$ gives rise to a dry spot or a dry-out hole in the center, which subsequently is subjected to the cooling process and eventually is frozen into the toroidal shape. For this case, the propagation of capillary wave (wavelength $\lambda= \gamma  /\rho V^2$) decays quickly and its travel distance ($l\sim \rho \lambda ^2 V/\eta$) is less than the characterized length scale of the droplet ($ l < R$), resulting in $We\,Ca\,> 1 \, (Ca = \eta U_0/\gamma)$ \cite{renardy2003pyramidal}. The physical mechanism of DTS will be addressed more comprehensively later (sketch in Fig. 2d).

\begin{figure}[b]
\center
 \includegraphics[scale=0.35]{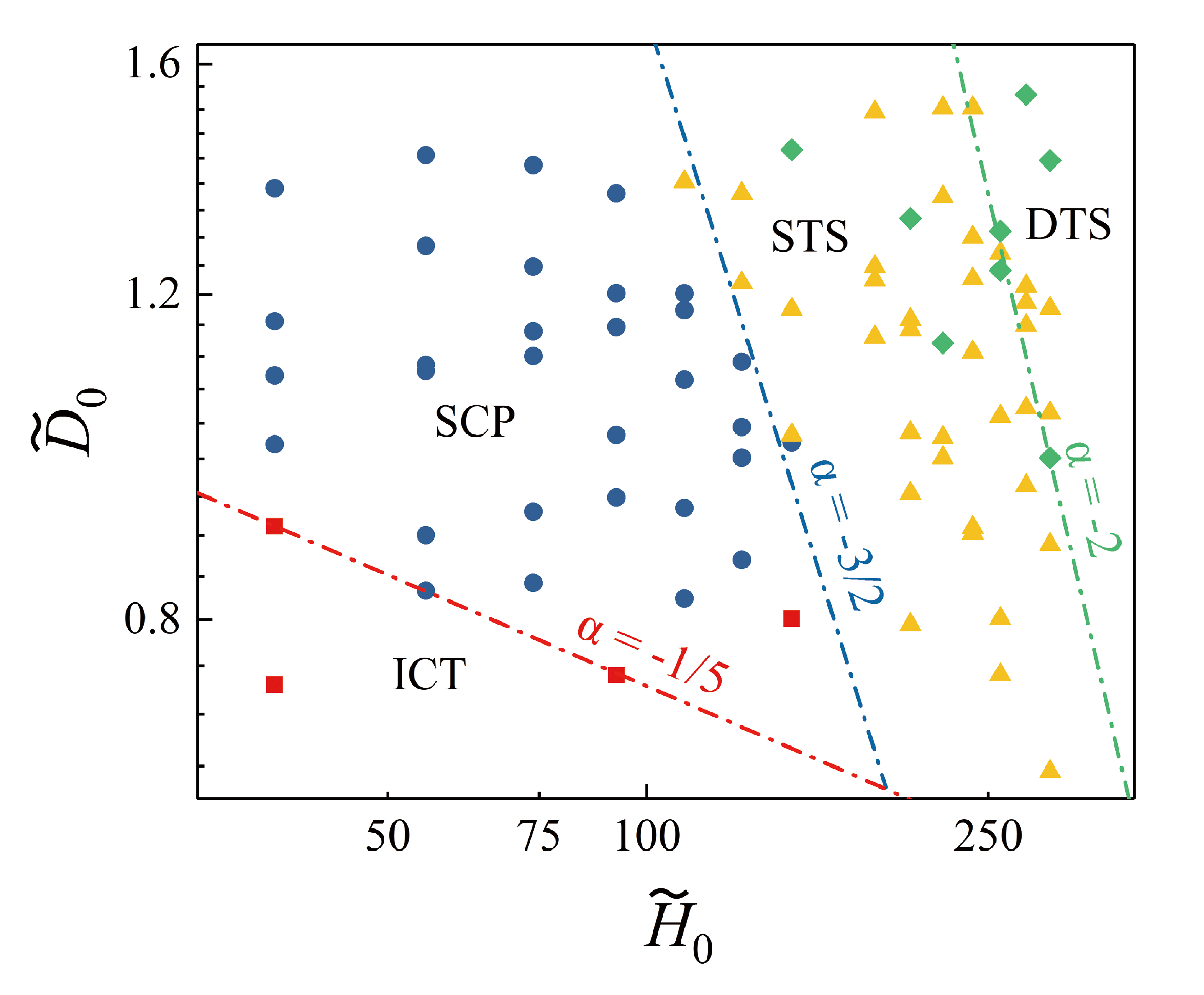}	
\caption{Phase diagram of frozen patterns dependent on $\tilde{H}_0$ and $\tilde{D}_0$ ($T_s = -20 \, \mathrm{^{\circ} C}$), and their transition described excellently by the scaling law of $\tilde{D}_0 \sim \tilde{H}_0^{\alpha}\,(\alpha = -1/5, -3/2, -2)$.}
     \label{fig_3}
\end{figure}

\begin{figure*}[t]
    \center
		  \includegraphics[scale=0.55]{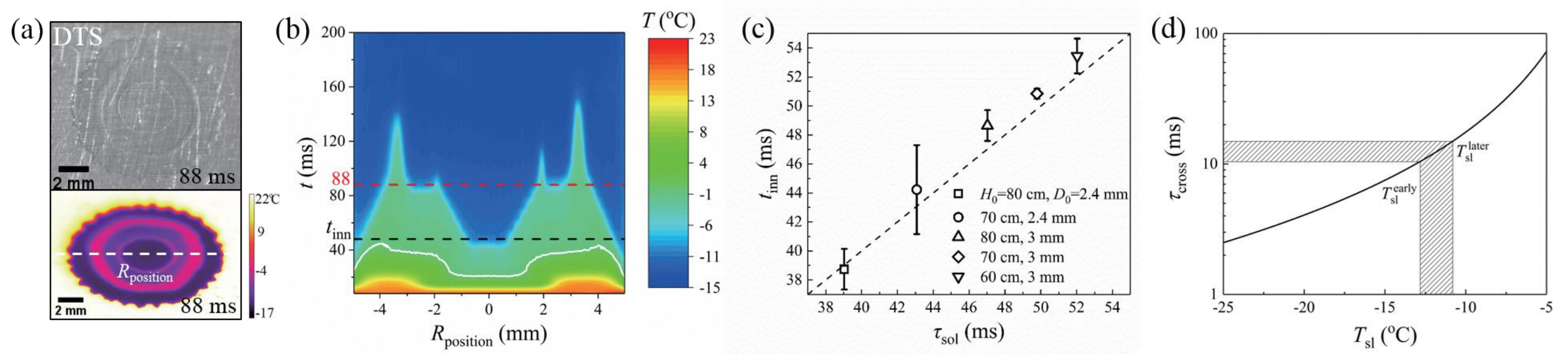}
	\caption{Mechanism of DTS due to the coupling of the droplet impacting hydrodynamics and solidification ($T_s = -15 \, \mathrm{^{\circ} C}$). (a) A typical synchronized high-speed image (top) and thermal image (bottom) at \emph{t} = 88 ms, revealing the correlation between the morphology and temperature distribution ($H_0 = 80$ cm). (b) Spatiotemporal evolution of the observed temperature along a horizontal line $R_\mathrm{position}$ in (a), the white line for the zero isotherm. (c) The identified $t_{inn}\simeq \tau_{sol}$, and error bars for three experiments. (d) The calculated $\tau_{cross}$ from Equation \eqref{equtimemodel} around 10 ms corresponding to $T_{sl}$ from the heat transfer model.}
\label{fig_4}
\end{figure*}

Furthermore, a scaling analysis of $\tilde{D}_0 \sim \tilde{H}_0^{\alpha}$ is performed to understand a phase diagram of these frozen patterns (Fig. 3a), which is obtained by controlling the droplet size ($D_0$) and falling height ($H_0$) at $T_s = -20 \,\mathrm{^{\circ} C}$ [$\tilde{D}_0 = D_0/k^{-1}, \tilde{H}_0 = H_0/k^{-1},  k^{-1} = (\gamma /\rho g)^{1/2}$ for the capillary length]. First, since SCP is characterized by the pinned edge and later unpinning from surface tension, the transition from ICT to SCP is by comparing $D_{max}$ [Equation \eqref{Dmaxscaling}] and the capillary length $k^{-1}$. Then, $D_{max} \sim k^{-1}$ leads to $\alpha = -1/5$,
\begin{equation}\label{scalingAtoB}
  \tilde{D}_{0}\sim \tilde{H}_0^{-1/5}.
\end{equation}

Secondly, as aforementioned mechanism of STS,  the transition criterion from SCP to STS is that $We\,Ca\ \sim 1$, resulting in $\alpha = -3/2$,
\begin{equation}\label{scalingBtoC}
  \tilde{D}_{0} \sim \tilde{H}_0^{-3/2}.
\end{equation}

Thirdly, the transition from STS to DTS is necessary to have a sufficiently faster timescale of solidification ($\tau_{sol}= h^2/D_s$) \cite{methodsolification} to better promote the ice formation for the observed inner holes or dry-out regions. The other relevant parameter is the timescale of the impacting process ($\tau_{imp}$), which is assumed to be the same as the timescale of contacting process for a bouncing drop independent on Weber number $\tau_{imp} = \tau_{cap}\sim (\rho D_{0}^{3}/\gamma)^{1/2}$ \cite{Richard2002contact}.  Then, $\tau_{cap}\sim \tau_{sol}$ gives rise to  $\alpha = -2$,
\begin{equation}\label{scalingCtoD}
  \tilde{D}_{0} \sim \tilde{H}_0^{-2}.
\end{equation}
Remarkably, this scaling law of $\tilde{D}_0 \sim \tilde{H}_0^{\alpha} \, (\alpha = -1/5, -3/2, -2)$ in Equation (\ref{scalingAtoB}-\ref{scalingCtoD}) for the pattern transition, has an excellent agreement with experiments, as shown by transition lines in Fig. 3.

Besides the droplet sizes and falling heights, these patterns can be manipulated by the substrate temperature. The phase diagram at $T_s = -10 \,\mathrm{^{\circ} C}$ and $-30\,\mathrm{^{\circ} C}$ (SM) not only demonstrates that patterns indeed are influenced dramatically by $T_s$, but also shows that DTS has more tendency to appear at a lower $T_s$, offering the possibility to achieve the desirable patterns through controlling $T_s$. Interestingly, despite the complicated effect of $T_s$, the simple scaling law for the pattern transition still holds for all these three phase diagrams.


Now we proceed to reveal the possible coupling dynamics of the droplet impacting and solidification responsible for the observed DTS. First, experimentally, by synchronizing a thermal camera (FLIR A 6750sc) with the fast-speed camera through an external trigger box (Video 1-4 in SM), the simultaneous snapshots (Fig. 4a) are both characterized by a common feature of two concentric rings, clearly demonstrating the correlation between the morphology and frozen process. The spatiotemporal evolution of temperature (Fig. 4b) further reveals this feature of two-concentric rings at \emph{t} = 88 ms (the red line) and the occurrence of the inner hole at $t_{inn}$ (the dark line) as well.

Secondly, the experimental $t_{inn}$ associated with the inner ice appearing in the DTS, is comparable with the calculated $\tau_{sol}$  \cite{methodsolification} for the central region of the liquid film to be transformed into ice,
\begin{equation}\label{timescaleinnsol}
t_{inn}\simeq \tau_{sol}.
\end{equation}
This identified timescale relationship is verified by more experiments for DTS produced at various parameters, such as $H_0$ and $D_0$ (Fig 4c), evidently showing the strong correlation between impacting dynamics and freezing process.

Thirdly, from a perspective of freezing kinetics, an ice front-advection model \cite{kant2020fast} is applied here to further understand DTS by considering the classical nucleation theory and the large-scale hydrodynamics on the droplet scale (sketch in Fig. 2d). During a cross-over timescale ($\tau_{cross}$), the ice size increases to be comparable with the thickness of the viscous boundary layer, and subsequently this advection of ice front at the substrate-liquid (water) interface should be related with the final frozen patterns.

From the classical nucleation theory, the growth rate of ice ($v_g$) is as below,
\begin{equation}
v_g (T_{sl}) =\frac{\textup{d}r}{\textup{d}t}= \frac{D}{\lambda }\left [ 1-\textup{exp}\left \{ -\frac{M\Delta G_{f,v}}{\rho N_{A}k_{B}T_{sl}} \right \} \right ],
\end{equation}
where $T_m = 0 \, \mathrm{^{\circ} C}$ is the melting temperature of water, $T_{sl}$ is the temperature at the substrate-liquid interface, and $\Delta G_{f,v} \sim (T_m-T_{sl})$ is the volumetric free-energy difference between solid and liquid \cite{methodnucleation}.  Then the ice size is $r_c = v_g t$ for a given $T_{sl}$, while the thickness of the viscous boundary layer grows as $\delta_v \sim \sqrt{\eta t/\rho}$ inside an impacting droplet \cite{roisman2009inertia}. The relationship between $\tau_{cross}$ and $T_{sl}$ is obtained from $r_c \sim \delta_v$ \cite{kant2020fast},
\begin{equation}\label{equtimemodel}
\tau_{cross} (T_{sl}) \sim \frac{\eta}{\rho v_g^{2}}.
\end{equation}
Based on simplified heating transfer models ($T_s= -15\, \mathrm{^{\circ} C}$) \cite{kant2020fast}(SM), $T_{sl}^{early} \approx -12.8 \, \mathrm{^{\circ} C}$ at the early stage by neglecting the latent heat from the heat diffusion model, while $T_{sl}^{later} \approx -10.8 \, \mathrm{^{\circ} C}$ at the later stage by including the released latent heat from a 1D two-phase Stefan model using Schwartz solution. According to the temperature zone between $T_{sl}^{early}$ and $T_{sl}^{later}$ (Fig. 4d), the estimated $\tau_\mathrm{cross} \sim 10\,$ ms is the same order of magnitude as the experimental $t_{inn}$.

In conclusion, we investigate frozen patterns for the water droplets impacting on a cold substrate through fast-speed images. These patterns can be manipulated by several physical parameters (the droplet size, falling height, and substrate temperature), and the scaling analysis has a remarkable agreement with the phase diagram. The observed double-concentric toroidal shape is attributed to the correlation between the impacting dynamics and freezing process, as confirmed by the spatiotemporal evolution of the droplet temperature, the identified timescale associated with the morphology and solidification ($t_{inn}\simeq \tau_{sol}$), and the ice front-advection model. These results for frozen patterns provide insight into the complex interplay of the rapid impacting hydrodynamics, the transient heat transfer, and the intricate solidification process.





\begin{thebibliography}{29}
\bibitem{methodnucleation}$M$ is the molar mass of water, $N_A$ is the Avegadro constant,
$k_B$ is the Boltzmann coefficient, $\lambda$ is the mean free path of a liquid molecule,
$D$ is the diffusion coefficient, $\Delta G_{f,v}=L_v(T_m-T_{sl})$ is
the volumetric free-energy difference between solid and liquid, and $L_v$ is the latent heat released per unit volume.
\end{thebibliography}
%

\end{document}